\documentclass[twocolumn,showpacs,aps]{revtex4}
\usepackage[dvips]{graphicx}
\usepackage{bm}
\begin{document}

\title{Waves on Icicles}
\author{Naohisa Ogawa and Yoshinori Furukawa
\footnote{ogawa@particle.sci.hokudai.ac.jp,~~frkw@lowtem.hokudai.ac.jp}}
\affiliation{Institute of Low Temperature Sciences, \\
Hokkaido University, Sapporo 060-0819 Japan}

\begin{abstract}
Icicles with wave patterns on their surfaces can sometimes be seen hanging
 from roofs of buildings. 
Surprisingly, most of these wave patterns are at intervals of about 1 cm. 
The reason for this uniformity of interval has not been clarified. 
Here we show a formula to explain this remarkable phenomenon 
by introducing a new instability theory. 
This theory is given by thermal diffusion in thin water layer streams 
flowing along the icicles. 
The streams change the temperature distribution and control waves of short wavelengths. 
The specific wavelength  (about 1 cm) can be determined 
by Laplace instability of the heat field 
in the atmosphere and by the thermal diffusion effect in thin-layer streams.
\end{abstract}
\maketitle

Icicles are often seen hanging from roofs of buildings in winter season in snowy regions. 
They are almost columnar in shape and their radii (or thickness) become smaller in the gravitational direction. 
The precise inner morphology of icicle is much more complicated and interesting like having hollows,
which is shown by Maeno et.al. \cite{1}.
But in this article we discuss only the surface instability of icicle as the periodically modulated thickness 
of an icicle along its longitudinal direction.
The intervals of wave (wavelength) appearing on icicles are almost uniform (about 1 cm), 
and do not depend on the length or thickness of the icicle or on the external temperature. 
Furthermore, such waves appear only on the surface of icicles that are enveloped 
by a layer of water flow of about 0.1 mm in thickness. 
The surface of a growing icicle changes from flat to wavy in a few hours \cite{matsu}.
Formation of such waves is also discussed qualitatively by Maeno et.al. \cite{1} as the conjecture.
But its scenario can not explain both the value of wavelength and its universality.
The mechanism to construct the wave here we show, is different from above one.

This kind of phenomenon is usually called surface instability. In this short article, 
we will discuss both the instability of the surface of an icicle and factors that determine 
the wavelength appearing on the surface of an icicle.
In 1997 Matsuda has given the experiment to construct the wavy ice \cite{matsu} in which 
the physical situation is the same as icicle.
He has made a gutter inclined at $\theta$ degrees and kept the static water flow on it. 
He put the apparatus in cold room with temperature -8 degrees centigrade.
Then the wavy ice has appeared under the water flow and its wave length is about 8 mm 
at $\theta= \pi/2$, which agrees with the case of icicle \cite{matsu}.
This means the waves appearing on icicle is induced by the interaction 
between thin water flow and thermal diffusion.

    In the study of crystal growth, surface instability is usually explained by Mullins-Sekerka's theory \cite{2}, 
which is based on two observations: Laplace instability and Gibbs-Thomson effect.
However, the mechanism underlying the instability that occurs on icicles cannot be explained by this theory. 
The reason for this is that the Gibbs-Thomson effect only applies to wavelength of about 10 micrometers; i.e., 
wavelength much shorter than those on icicles (about 1 cm). 
Furthermore, Laplace instability does not occur in a thin liquid layer; it only occurs when the thickness of the layer is wider
 than the wavelength. Thus, Laplace instability could only occur on icicles if the thickness of liquid layer 
 were larger than the wavelength of 1 cm, which is never the case.

    Let us then consider the characteristics of a fluid. 
There are various instability theories in fluid mechanics. 
Benney's liquid film has often been used to study the characteristics of 
a thin water layer \cite{3}. 
This theory explains why waves appear on the surface of thin water flow 
on a sloping road on a rainy day. 
This phenomenon is caused by gravity, viscosity and surface tension. 
However,these are not static waves but moving waves. 
In the case of icicles, waves move down an icicle two-times 
faster than the average velocity of a fluid. 
Thus, the formation of waves on an icicle can not be explained only on the basis 
of characteristics of such hydrodynamical waves. 

    We therefore consider not only the fluid characteristics 
which flows along an icicle  but also the thermal diffusion effect to express the static wave on the surface of icicle.
Our program is the following. First, we solve the hydrodynamics of thin-layer fluid, 
which is not used in the usual Mullins-Sekerka theory. 
Next, we consider the thermal diffusion in thin-layer fluid and in the air surrounding an icicle. 
The thermal diffusion in air is very important 
since Laplace instability cannot occur in a water layer. As a boundary condition, 
the heat flow should be conserved on the boundary between air and the water. 
The thermal properties in water fluid and in air give all the necessary information 
on the formation of surface of icicles.

    Let us start with the case of an icicle having a boundary surface (ice-water) 
with a small perturbation. 
We add a perturbation to the radius $R$ of an infinitesimally long  columnar ice (model of icicle),

\begin{equation}
 R \to R + \delta(t) \sin kx, 
\end{equation}
(x-axis is along the length of the icicle) 
and we consider the time development of $\delta(t)$. 
We define the y-axis as perpendicular to the ice surface, and use the variable 
$y= r - R$ with $\mid y \mid << R$.
The thin water layer on this surface has mean thickness $h_0 \sim 0.1mm$.
For considering such a film fluid, it will be easier to have a fluid dynamical model as 
thin water layer on a inclined gutter just like Matsuda's experiment \cite{matsu}.
(x-axis is taken along slope, and y-axis is taken normal to slope surface.)
This simplification is well established since $h_0 \ll R$.
This situation for fluid mechanics is usually discussed for flat gutter,
and shown the specific surface velocity
\begin{equation}
U = \frac{gh_0^2}{2\nu},
\end{equation}
with parabolic velocity distribution,
\begin{equation}
v_x = U(2\frac{y}{h_0} - (\frac{y}{h_0})^2),
\end{equation}
where $\nu$ is the viscosity ($1.8 \times 10^{-6} m^2/s$) and $g$ is the gravitational acceleration.
(For this calculation, see for e.g. the exercise in Landau's text book \cite{3}.)
If we observe mean surface velocity $U$, we obtain $h_0$ 
and flow quantity ${\bf Q}=2\pi R \bar{U} h_0$,
where $\bar{U}= \frac{1}{h_0} \int_0^{h_0} v_x dy = 2U/3$ is the mean velocity.
In Matsuda's experiment \cite{matsu}, he used the flow ${\bf Q}=160 ml/hr$ and width of gutter 
$W= 3cm$ ($W$ corresponds to $2\pi R$), where the wave appears most clear.
This gives $\bar{U} h_0 = 1.48 \times 10^{-6} m^2/s$. From $\bar{U}= 2U/3$ and $U = \frac{gh_0^2}{2\nu}$, 
we get $U=2.4 \times 10^{-2} m/s$ with $h_0 = 0.93 \times 10^{-4} m$. 
On the other hand, his direct observation of surface mean velocity by using test fine grain gives $U=4 \times 10^{-2} m/s$.
So we use $U= (2.4 \sim 4) \times 10^{-2} m/s$ with $h_0 = (0.93 \sim 1.21) \times 10^{-4} m$ 
as the experimental surface velocity and thickness of water layer.

Our problem with small modulation $\delta \sin kx$ on boundary can be expressed as perturbation to above solution.
To do this we make nondimensional coordinate $x_*$ and $y_*$ as
$$ x_* = kx, ~~~~~ y_* = y/h_0,$$
and we put off * in a while.
In this notation solid surface is given as $y=\eta \sin x$, where $\eta= \delta / h_0$.
The fluid in such a thin-layer has boundary conditions as follows.
The fluid velocity equals to zero at the ice-water boundary due to viscosity, 
and the gradient of fluid velocity equals to zero on the air-water surface due to the vanishing of shearing stress.

By solving static Navier Stokes equation \cite{4} with above mentioned boundary conditions 
and with long wavelength approximation $\mu \equiv kh_0 \sim 10^{-2}$, 
we obtain the hight of air-water boundary surface in nondimensional form given as,
\begin{equation}
y=1 + \eta \sin x ,
\end{equation}
which means the thickness of water layer is uniform at lowest order of long wavelength approximation.
(In first order approximation, we have modulation in its thickness 
but it is not thinner at protrusions. The thickness modulation is like $\cos kx$ 
against the icicle modulation: $\sin kx$. But these effects are irrelevant. \cite{4})
The stream function of water is obtained as
\begin{equation}
\psi_* = -\frac{1}{3}(y-\eta \sin x)^3 + (y-\eta \sin x)^2,
\end{equation}
where $\psi_*$ is also the non-dimensional stream function as,
$$\psi = h_0 U \psi_*,$$
and we put off * again.
 The form of such stream function is not surprising, and it means just the parabolic form of velocity fields, where the velocity vanishes at modulated solid-water boundary.

The next task is to solve the thermal diffusion equation with a flow background as,
\begin{equation}
 \triangle T - \frac{\vec{v}}{D} \cdot \vec{\nabla} T = \frac{1}{D}
\frac{\partial T}{\partial t},
 \end{equation}
where $D \equiv \frac{\kappa}{\rho c} \sim 1.3 \times 10^{-7} m^2/s$ is the thermal diffusion constant
 and we used non-compressive condition.
We rewrite above form by using  stream function $\psi$,  
neglecting higher order in parameter $\mu \equiv k h_0$ as long wavelength approximation, 
and under the static thermal distribution condition.
(All variables are written in dimensionless form except temperature.)
\begin{equation}
\frac{\partial^2 T}{\partial y^2} = \alpha (\frac{\partial T}{\partial x}\frac{\partial \psi}{\partial y} 
- \frac{\partial T}{\partial y}\frac{\partial \psi}{\partial x}).
\end{equation}

Note that the dimensionless parameter $\alpha \equiv \mu U h_0/D$ 
is proportional to wave number k.  
The value of $\alpha$ is about 2  experimentally. 
By using boundary condition on solid-water surface ($y= \eta \sin x$)as
\begin{eqnarray}
T(SW) &=& T_M = const.,\\
Q(SW) &=& -\kappa \frac{\partial T}{\partial y}\mid_{y= \eta \sin x}\equiv -\kappa a(x),
\end{eqnarray}
we obtain the quantities on air-water surface \\
($y= 1+ \eta \sin x$) by solving differential equation.
\begin{eqnarray}
T(AW) &=& T_M + [1 + \frac{7}{60} \hat{D} + \frac{13}{3360}\hat{D}^2] a(x) ,\label{T1} \\
Q(AW) &=& -\kappa[1 + \frac{5}{12} \hat{D} +\frac{239}{10080} \hat{D}^2] a(x),\label{Q1}
\end{eqnarray}
where $T_M$ is the melting point, $Q$'s are heat flows, $a(x)$ is an unknown gradient of temperature on solid surface, and $\hat{D}$ means $\alpha d/dx$.The higher order terms of $\alpha$ are negligible since their coefficients are effectively small.

It should be noted that the temperature of the air-water surface is 
different from the one in the environment. 
The temperature in atmosphere changes greatly near the surface of an icicle. 
Thus, we must also consider the thermal diffusion in air. 
Since we consider the static diffusion, our equation becomes Laplace equation. 
We write it in cylindrical coordinate as
\begin{equation}
[\frac{\partial^2}{\partial r^2}+ \frac{1}{r}\frac{\partial}{\partial r}+
\frac{1}{r^2}\frac{\partial^2}{\partial \theta^2}+
\frac{\partial^2}{\partial x^2}]T(r,\theta, x) =0.
\end{equation}
Note that our previous coordinate $y$ is related to $r$ as 
$$ r = R + y. $$
We suppose the existence of axial symmetry, and we 
take $\partial T/ \partial \theta =0$.
Therefore we work with
\begin{equation}
[\frac{\partial^2}{\partial r^2}+ \frac{1}{r}\frac{\partial}{\partial r}+
\frac{\partial^2}{\partial x^2}]T(r, x) =0.
\end{equation}
Since we have oscillating surface in x-direction, 
it is natural to construct the solution in the form,
\begin{equation}
T(r,x) = f(r) + g(r)\sin (kx + \phi),
\end{equation}
where $f(r)$ satisfies
\begin{equation}
[\frac{d^2}{d r^2}+ \frac{1}{r}\frac{d}{d r}]f(r) =0,
\end{equation}
and $g(r)$ satisfies
\begin{equation}
[\frac{d^2}{d r^2}+ \frac{1}{r}
\frac{d}{d r}-k^2]g(r) =0.
\end{equation}

Then the solution is given by using dimensional variables $x,\tilde{y} \equiv y-h_0, \tilde{R} \equiv R+h_0$ as
\begin{eqnarray}
T(x,y) &=& A + B \log(1+\tilde{y}/\tilde{R}) \nonumber \\
&&  ~~~~~ + C K_0(k(\tilde{R}+\tilde{y})) \sin (kx + \phi),
\end{eqnarray}
where A,B, and C are the coefficients and $K_0$ is the modified Bessel function.
Last term is induced from boundary condition at water surface.
The reader might think  the appearance of logarithmic term strange, since it diverges at large $r$.
But the appearance of such a term is natural for infinitesimally long axially symmetric source.
In our system the length of icicle is finite. Therefore this solution is valid just around the icicle, 
and far from icicle source we must treat the source just like point and we need to connect two solutions at $r \sim L$ 
(L is the order of length of icicle). In this way the coefficients A,B and C will be determined.
But we have another physical observable, such as, mean growth rate of icicle which determine coefficient B.
The information of temperature at infinity is included into this observable.

At $\tilde{y}=\delta \sin kx$ (AW surface), temperature and thermal flow are given by above solution by using taylor expansion.
\begin{eqnarray}
T(AW) &=& A + [C K_0(k\tilde{R})\cos \phi -\frac{LV}{\kappa_0}\delta] \sin kx \nonumber \\
&&+ C K_0(k\tilde{R})\sin \phi \cos kx, \label{T2}\\
Q(AW)&=& LV - [\kappa_0 C k K'_0(k\tilde{R}) \cos \phi + \frac{LV \delta}{\tilde{R}}] \sin kx 
\nonumber \\
&&- \kappa_0 C k K'_0(k\tilde{R}) \sin \phi \cos kx, \label{Q2}
\end{eqnarray}
$V$ is the mean growth rate of icicle defined by 
$$V \equiv  <Q>/L,$$
with $L$ the latent heat, and $<~>$ means the average along x-direction.
This value determines coefficient B as,
$$B = -\frac{L\tilde{R}V}{\kappa_0}.$$
$\kappa_0$ is the thermal conductivity of air.
We treat $C$ as order $\delta$ and we took up to first order in $\delta$.

Note that the thermal-diffusion inside an icicle is not considered. 
Since the surface of icicle is the same temperature as melting point, 
the temperature inside an icicle can be considered as uniform.
We now give continuity condition of temperature and heat flow conservation 
on the water-air boundary surface. 
By replacing (\ref{T1}) and (\ref{Q1}) in dimensional form, 
and compare with (\ref{T2}) and (\ref{Q2}),
all the coefficients are determined as well as the function $a(x)$. 
The growth rate of solid is determined from this function as 
\begin{equation}
v(x) = -\kappa a(x)/L = V + b \sin kx + c \cos kx,
\end{equation}
where b and c are obtained some coefficients.
We can obtain the amplification rate of fluctuation by $\dot{\delta}/\delta = b/\delta$.
We have two mixed effects for the growth rate of fluctuation. 
One is the diffusion property of temperature in the atmosphere. 
The other is the effect of fluid for thermal diffusion in thin-layer. 
In total, we obtain the amplification rate of fluctuation under the condition $kR >>1/2$.
\begin{equation}
\frac{\dot{\delta}}{\delta} = Vk \times 
\frac{1-\frac{239}{10080}\alpha^2}{1+\frac{1272}{10080}\alpha^2 +(\frac{239}{10080})^2 \alpha^4},
\end{equation}

where $\alpha \propto k$ , and $V$ is the mean growth rate.  
This equation is interpreted as 
\begin{flushleft}
R.H.S = (Laplace instability) 
\end{flushleft}
\begin{center}
  ~~~~$\times$ (fluid effect for thermal diffusion).
\end{center}
   Therefore, thermal diffusion in atmosphere enhances fluctuation of shorter wavelength,  
but fluid effect for thermal diffusion tries to suppress such fluctuation of shorter wavelength. 
From these two effects, amplification rate has a peak value. 
The maximum amplification factor is given at definite wave number.
$$\alpha_{max} \sim 2.2.$$
The corresponding wavelength is obtained by using the definition of $\alpha$ as,
$$ \lambda = \frac{2\pi U h_0^2}{D \alpha}.$$
The experimental data: 
$$U= (2.4 \sim 4) \times 10^{-2} m/s,~~~h_0 = (0.93 \sim 1.21) \times 10^{-4} m$$
with $\alpha_{max} \sim 2.2.$ gives $\lambda = 5mm \sim 13mm$, which well agrees with observed wave length 8mm.

Let us discuss our obtained result by comparing with the discussion given by Maeno.et.al.\cite{1}.
Our obtained result comes from two important effects.
One is the thermal diffusion in air and its Laplace instability, 
another one is the effect of fluid for thermal diffusion in thin water.
The fluid effect shuffles the static temperature distribution and so the higher frequency wave is suppressed.
On the other hand Maeno et.al. \cite{1} have given the conjecture on the formation of ribs on icicle.
This is from three points as follows.

1: If we had an protrusion accidentally, the water flow on it may be thinner and cooled much faster.
2: The protrusions are more exposed in the surrounding cold air and giving rapid growth of ribs.
3: Since the water flow have larger viscosity, the water tends to be stagnant just upstream of ribs and giving slower growth of icicle.

The first and third one are irrelevant in long wavelength approximation which we used, 
and the thickness of water layer is almost uniform in our explicit fluid mechanical calculation.
This will be well understood by the difference between thickness 0.1mm and wavelength 10mm.
The second one may correspond to the Laplace instability in air which we used.
But this is not enough to express the wave properties. 
Really all of above three remarks enhance shorter wavelength fluctuations, 
and do not determine specific wavelength.
So we have used not only the property of water thickness, 
but also the velocity distribution of water flow which affects thermal diffusion.
This effect breaks the Laplace instability at shorter wavelength, 
and we obtain peak growth rate at definite wavelength as the quantitatively good result.

The wavelength depends on the fluid flow \\
${\bf Q}\equiv 2\pi R \bar{U} h_0$ on icicle surface:
\begin{equation}
\lambda_{\max} = \frac{2\pi U h_0^2}{D \alpha_{max}} 
= \frac{1}{D \alpha_{max}}(\frac{\nu}{g \pi})^{1/3}(\frac{3{\bf Q}}{2R})^{4/3}, \label{sol}
\end{equation}
where $R$ is the mean radius of the icicle. 
The temperature of the environment changes the mean growth rate of solidification, 
but wavelength is independent of it. The wavelength depends on the ratio of ${\bf Q}$ to $R$ but not only on ${\bf Q}$. 
We then have a universal wavelength by assuming that ${\bf Q}$ is proportional to $R$. 
The fluid effect that suppresses short wavelength fluctuation is somewhat similar to the Gibbs-Thomson effect. 
But their origins are quite different, and they work in different length scales. 
Further experimental test will be  necessary to check the conclude validity of this theory by using (\ref{sol}).
But here we have shown how the wave can be presented on icicle as instability theory.
Our proposed instability theory is quite new and may be applied to a wide range of phenomena. 
For example, the similar phenomenon occurring on stalactites can be explained in the same way 
by changing thermal diffusion to solution diffusion.

\begin{acknowledgments}
The authors are grateful to Prof. R. Takaki for his valuable discussion. 
One of the authors N.O. thanks Prof. K. Fujii for his encouragement during this study. 
The authors are also grateful to Prof. E. Yokoyama, Dr. Nishimura, and Prof. R. Kobayashi 
for their helpful discussions and encouragement and to Ms. K. Norisue for her help in English corrections.
\end{acknowledgments}

\end{document}